\documentclass[preprint,showkeys,superscriptaddress,nofootinbib]{revtex4}

\usepackage[dvips]{graphicx,color}
\usepackage{epsfig}
\usepackage{amsmath}
\usepackage{amssymb}
\usepackage{booktabs}
\usepackage{url}
\usepackage[dvipdfm,bookmarks=true,bookmarksnumbered=true,bookmarkstype=toc]{hyperref} 

\allowdisplaybreaks[4]
\newcommand{\lsim}{\raise0.3ex\hbox{$\;<$\kern-0.75em\raise-1.1ex\hbox{$\sim\;$}}}
\newcommand{\gsim}{\raise0.3ex\hbox{$\;>$\kern-0.75em\raise-1.1ex\hbox{$\sim\;$}}}

\begin{document}


\title{Correlation between flavour violating decay of long-lived slepton and tau  
in the coannihilation scenario with Seesaw mechanism}

\author{Satoru Kaneko}
\email{satoru@cftp.ist.utl.pt}
\affiliation{Juntendo University, Graduate School of Medicine,
Hongo 2-1-1, Bunkyo-ku, Tokyo, 113-8421, Japan}

\author{Hiroki Saito}
\email{hiroki@phy.saitama-u.ac.jp}
\affiliation{Department of Physics, Saitama University, 
Shimo-Okubo, Sakura-ku, Saitama, 338-8570, Japan}

\author{Joe Sato}
\email{joe@phy.saitama-u.ac.jp}
\affiliation{Department of Physics, Saitama University, 
Shimo-Okubo, Sakura-ku, Saitama, 338-8570, Japan}

\author{Takashi Shimomura}
\email{stakashi@yukawa.kyoto-u.ac.jp}
\affiliation{Yukawa Institute for Theoretical Physics, 
Kyoto University, Kyoto 606-8502, Japan. 
}

\author{Oscar Vives}
\email{oscar.vives@uv.es}
\affiliation{Departament de F\'isica Te\`orica 
and IFIC, Universitat de Val\`encia - CSIC, 
E46100, Burjassot, Val\`encia, Spain}

\author{Masato Yamanaka}
\email{yamanaka@cc.kyoto-su.ac.jp}
\affiliation{Maskawa Institute, 
Kyoto Sangyo University,  
Kyoto 603-8555, Japan. 
}

\preprint{YITP-11-16,~FTUV-11/0402,~IFIC/11-06,~STUPP-11-207,~MISC-2011-02}

\date{\today}

\begin{abstract}
We investigate flavour violating decays of the long-lived lightest
slepton and the tau lepton in the coannihilation region of the Minimal
Supersymmetric Standard Model with a Seesaw mechanism to generate
neutrino masses.  We consider a situation where the mass difference
between the lightest neutralino, as the Lightest Supersymmetric particle
(LSP), and the lightest slepton, as the Next-to-LSP, is smaller than the
mass of tau lepton. In this situation, the lifetime of the lightest
slepton is very long and it is determined by lepton flavour violating (LFV) couplings
because the slepton mainly consists of the lighter stau and the
flavour conserving 2-body decay is kinematically forbidden. 
We show that  the lifetime can change many orders of magnitude 
by varying the Yukawa couplings entering the Seesaw mechanism.
We also show that branching ratio of LFV tau decays are strongly
correlated with the lightest slepton lifetime. Therefore the branching
ratios of LFV tau decays can be determined or constrained by measuring
the slepton lifetime at the LHC experiment.
\end{abstract}

\maketitle

\section{Introduction}\label{sec:introduction}

Neutrino oscillation experiments \cite{Fukuda:2002pe, Ahmed:2003kj,
  Araki:2004mb, Ahn:2006zza} have confirmed that neutrinos are massive
and mix with each other. Thus, flavour is violated in the
lepton sector similarly to the quark sector flavour violation
described by the CKM mixing matrix. These experimental results require
that, today, the Standard Model (SM) of particle physics must include a way to
accommodate neutrino masses and mixing.
In this extension of the SM, Lepton Flavour Violation (LFV) in charged lepton sector should also occur through 
neutrino mixing in loop corrections, although it has 
not been found so far. In fact, the rates of LFV of charged
leptons induced by neutrino mixing, being proportional to the mass differences of neutrinos, 
are far below the present and near-future experimental sensitivities.
Therefore, the discovery of flavour violating processes in the charged lepton 
sector would be a clear evidence of new physics beyond the SM.

One of the most attractive mechanisms to realize the observed tiny neutrino 
masses is 
the so-called Seesaw mechanism~\cite{Minkowski:1977sc, Yanagida:1979, 
Gell-mann:1979, Mohapatra:1979ia,Schechter:1980gr}, where Majorana 
right-handed neutrinos with heavy masses are introduced. Due to the 
presence of these right-handed neutrinos, a Yukawa coupling and a Majorana 
mass term for the right-handed neutrinos are allowed, and left-handed 
neutrinos acquire masses  through an effective dimension-five operator, after the Higgs scalar develops a vacuum expectation 
value. 
Then, the masses of left-handed neutrinos are a function of the neutrino Yukawa couplings and the heavy right-handed Majorana masses 
and they become naturally tiny because they are suppressed  by these heavy Majorana masses. Therefore, the observed light neutrino 
masses and mixing provide information on the heavy right-handed neutrino masses and the neutrino Yukawa couplings in the Seesaw scenario. 
It was pointed out in \cite{Casas:2001sr}, however, that apart from the unknown right-handed Majorana masses
there exists a 
complex orthogonal matrix in the parametrization of the neutrino Yukawa 
coupling,  whose six parameters can not be determined by low-energy neutrino oscillation experiments.
In a supersymmetric (SUSY) extension of the 
Seesaw mechanism, flavour mixing among sleptons (scalar partners of 
leptons) are induced at low scale from the neutrino Yukawa coupling through 
Renormalization Group Equations (RGEs)~\cite{Borzumati:1986qx, Hisano:1995nq,
Hisano:1995cp}.
 Since flavour violating decays and flavour conversions of leptons 
occur via the slepton mixing~\cite{Borzumati:1986qx, Hisano:1995nq,Hisano:1995cp}, 
the parameters can be determined or constrained by 
measurements 
of these processes. A lot of theoretical works have been done for this purpose 
(see some recent works~\cite{Raidal:2008jk,Hisano:1998fj,Masina:2002mv,Petcov:2003zb,Lavignac:2001vp,Masiero:2002jn,Lavignac:2002gf,Ellis:2002fe,Davidson:2003cq,Davidson:2001zk,Calibbi:2008qt}), and several experiments to explore LFV processes are 
ongoing or will start in the near future~\cite{Cui:2009zz, PRISM, Mu2e, Adam:2009ci,O'Leary:2010af,Akeroyd:2004mj,Marciano:2008zz}.

Apart from the LFV processes of leptons, it is clear that once we are able
to produce sleptons in colliders, their decays can also provide
information on the flavour violating entries in the slepton mass
matrix. In particular, it was shown in \cite{Kaneko:2008re}, that in a scenario of the Minimal Supersymmetric SM (MSSM) 
where the lightest slepton and the lightest neutralino are nearly degenerate, as it happens in part of the
coannihilation region~\cite{Griest:1990kh}, the
lifetime of the lightest slepton has a very good sensitivity to small LFV
parameters.
In this scenario, the Lightest Supersymmetric Particle (LSP) is the lightest 
neutralino which is almost pure Bino, and the Next-to-LSP (NLSP) is the 
lightest slepton, which mainly 
consists of the right-handed stau. 
When the mass difference between the LSP neutralino and the NLSP slepton is 
smaller than the tau mass, the decay of the NLSP slepton 
into tau and neutralino is kinematically forbidden.
Then, flavour conserving decays are only $3$-body (or $4$-body) decays 
into a tau neutrino, a pion and the neutralino (a tau neutrino, two leptons 
and the neutralino). The decay widths of flavour conserving processes are highly 
suppressed due to the additional Fermi coupling and the small phase space  
, and therefore the lightest slepton becomes long-lived \cite{Jittoh:2005pq}.
On the other hand, flavour violating $2$-body decays into an electron
(or a muon) and the neutralino are kinematically allowed. They become 
dominant if the suppression due to LFV couplings is looser than the
suppression of the flavour conserving decays. In this situation,
the lifetime of the slepton is determined by LFV parameters and is
sensitive to very small values of the LFV parameters. 
In Ref.~\cite{Kaneko:2008re}, it was shown that such a small mass
difference is realized in the Constrained MSSM (CMSSM) consistent with present
constraints from terrestrial experiments and cosmological
observations.
It was also discussed that determination of the LFV parameters via the lifetime 
could be possible in the ATLAS detector~\cite{Kaneko:2008re} when the 
sleptons decay inside the detector. Therefore, the LHC experiment 
provides a good opportunity to explore the LFV of  the sleptons in this nearly degenerate scenario.

In this work, we will analyze this scenario of nearly degenerate LSP and NLSP
in the MSSM with the Seesaw mechanism. We assume that slepton mass
matrices are perfectly universal at the Grand Unification Theory (GUT) scale 
and flavour violating entries in these matrices are generated by RGE
evolution in the presence of the neutrino Yukawa coupling. Then, the
lifetime of the slepton is mainly determined by the mixing of
left-handed sleptons, because larger flavour mixing is induced to the
left-handed sleptons than to the right-handed sleptons. Similarly, in
this scenario, the rates of flavour violating tau decays are strongly
related with the lifetime of the slepton because the NLSP slepton
mostly consists of the stau. Thus, it is worthwhile to study
correlations between the lifetime of the slepton and the branching
ratios of the LFV tau decays to obtain information on the parameters
of the Seesaw mechanism.

The rest of the paper is organized as follows. In
Sec.~\ref{sec:setup:-mssm-with-RN}, we briefly review the CMSSM with the
Seesaw mechanism. In Sec.~\ref{sec:long-lived-slepton}, we give  
expressions for the decay rates of the slepton and LFV tau decays, and then 
derive relations among branching ratios and lifetimes. 
The results of our numerical calculation are
shown in Sec.~\ref{sec:numerical-result}. Finally, we summarize and
discuss our results in Sec.~\ref{sec:conclusion}.

\section{The CMSSM with right-handed neutrinos} 
\label{sec:setup:-mssm-with-RN}

We start our discussion with a brief review of the Seesaw mechanism in the CMSSM. The CMSSM is 
defined at the GUT scale, 
$M_{\mathrm{GUT}} \simeq 2 \times 10^{16}$ GeV, by four parameters and a sign:
\begin{eqnarray}
\left\{M_{1/2},\ m_0,\  A_0,\ \tan\beta,\ {\rm sign}(\mu)\right\},
\end{eqnarray}
where $M_{1/2}$ and $m_0$ are the universal gaugino and scalar masses,
respectively, and $A_0$ is the universal trilinear coupling.
$\tan\beta$ is the ratio of the vacuum expectation values of up-type
Higgs ($v_u$) to down-type Higgs ($v_d$), $\tan\beta=v_u/v_d$, and
$\mu$ is the Higgs mass parameter in the superpotential. It is
important to emphasize that in this model, soft SUSY breaking terms
are completely universal at the GUT scale, i.e. soft mass matrices are
proportional to the identity matrix and trilinear couplings are
proportional to the Yukawa couplings. This universality is broken in
the quark sector by radiative corrections due to the presence of the up- and
down-type quark Yukawa couplings, but it is preserved in the lepton sector if
neutrinos are massless.
When Majorana right-handed neutrinos are introduced in the CMSSM, a 
neutrino Yukawa coupling and a Majorana mass term for the right-handed 
neutrinos are allowed in the superpotential. 
In the basis where the charged lepton Yukawa coupling and the right-handed 
neutrino mass matrix are diagonal\footnote{Without loss of generality, we 
can always take this basis.}, the leptonic part of the superpotential is given 
by
\begin{eqnarray}
\mathcal{W}=
(\hat{y}_\ell)_{\alpha} L_{\alpha} H_d E_{\alpha}^c+
(y_\nu)_{\alpha i} L_{\alpha} H_u N_i^c+
(\hat{M}_{R})_{i}N_i^c  N_i^c + h.c.,
\end{eqnarray}
where $i=1$-$3$ and $\alpha = e,\mu,\tau$. The charged lepton Yukawa coupling, 
the neutrino Yukawa coupling and the diagonal right-handed neutrino mass 
matrix are denoted as $\hat{y}_\ell$, $y_\nu$ and $\hat{M}_R$ where $\hat{M}_R={\rm diag}(M_{R1},M_{R2},M_{R3})$, 
respectively. 
The superfields corresponding to the left-handed leptons, right-handed
charged leptons and right-handed neutrinos are denoted as
$L_{\alpha}$, $E^c_{\alpha}$, and $N^c_i$.  The superfield of the up-type (the down-type) 
Higgs doublet is as $H_{u(d)}$. Throughout this paper, we use Greek indices to denote 
flavour eigenstates (= mass
eigenstate of charged leptons) and Latin indices to mass eigenstates of
both of left- and right-handed neutrinos. After the electroweak
symmetry breaking, the effective mass matrix of the left-handed
neutrinos, $m_\nu$, is given by
\begin{eqnarray}
m_\nu=-\frac{v_u^2}{2}\,y_\nu\cdot \hat{M}_R^{-1}\cdot y_\nu^{T}.
\label{seesaw}
\label{nmass}
\end{eqnarray}
From this equation, we can see that neutrino masses are suppressed by a factor $(v_u^2/M_{Ri})$, and hence tiny masses are naturally generated. 
The mass matrix of the neutrinos, Eq.~(\ref{nmass}), is a complex symmetric matrix and is 
diagonalized as
\begin{eqnarray}
   \hat{m}_\nu 
   = U^T\cdot m_\nu\cdot U 
   = {\rm diag}(m_1,m_2,m_3),
\end{eqnarray} 
where $U$ is the neutrino mixing matrix (or the MNS matrix~\cite{Maki:1962mu}) 
that contains three mixing angles, one Dirac-type and two 
Majorana-type CP violating phases. The diagonal matrix, $\hat{m}_\nu$, 
contains three light neutrino masses, $m_i$. 
By inverting the seesaw relation Eq.~(\ref{seesaw}), we obtain the following 
expression for the neutrino Yukawa coupling~\cite{Casas:2001sr} 
\begin{eqnarray}
   y_\nu 
   = \frac{\sqrt{2}i}{v_u} U^{\ast} 
   \cdot \sqrt{\hat{m}_\nu} \cdot W 
   \cdot \sqrt{\hat{M}_R}~,  \label{eq:7}
\end{eqnarray}
where $W$ is a general complex orthogonal matrix. 
Notice that this orthogonal matrix, $W$, does not contribute to the left-handed neutrino 
masses and mixing, and hence can not be measured in neutrino oscillation experiments. We can parameterize $W$ as 
\begin{align}
 W=
 \begin{pmatrix}
  \cos\omega_1 & -\sin\omega_1 & 0 \\
  \sin\omega_1 & \cos\omega_1  & 0 \\
  0            & 0             & 1
 \end{pmatrix}
 \begin{pmatrix}
  \cos\omega_2 &  0 &\sin\omega_2   \\
  0            &  1 & 0 \\
  -\sin\omega_2 & 0 &\cos\omega_2  
 \end{pmatrix}
 \begin{pmatrix}
  \cos\omega_3 & -\sin\omega_3 & 0 \\
  \sin\omega_3 & \cos\omega_3  & 0 \\
  0            & 0             & 1
 \end{pmatrix},
 \label{eq:99}
\end{align}
where, in principle, the angles $\omega_{1,2,3}$ are complex numbers. 
In fact, the different elements of the Yukawa coupling, Eq.~(\ref{eq:7}), 
can change many orders of magnitude due to the imaginary parts of these 
complex angles, even if all other parameters are fixed. However, as we will see later, these complex angles 
affect LFV processes, and large values of elements would violate the present bounds. By this reason,  in this paper,
we will take $ -\frac{3}{2}\pi < \mathrm{Im}(\omega) <  \frac{3}{2}\pi$ for numerical analysis in Sec.\ref{sec:numerical-result}.

After introducing the neutrino Yukawa coupling and right-handed Majorana masses 
in the superpotential, they modify the slepton mass matrix at the loop level and generate 
lepton flavour violating entries through RGE running in analogy to the quark sector.
These flavour violating off-diagonal elements induced in the slepton mass 
matrix and in the slepton trilinear coupling can be estimated at one-loop
in the leading-log approximation as~\cite{Borzumati:1986qx, Hisano:1995cp,Petcov:2003zb}: 
\begin{eqnarray}
   \Delta M_{LL \alpha \beta}^{e 2}
   &\simeq& 
   -\frac{1}{8\pi^2} 
   \left(3m_0^2+A_0^2\right) 
   \left(y_\nu^{\dag}L\, y_\nu\right)_{\alpha \beta}
   ,\ \ \ 
   L_{ij} 
   \equiv 
   \log \left(\frac{M_{GUT}}{M_{Ri}}\right) 
   \delta_{ij}, \label{eq:4} 
   \\
   \Delta M_{RR \alpha \beta}^{e 2}
   &\simeq& 
   0, \label{eq:5} 
   \\
   \Delta M_{LR \alpha \beta}^{e}
   &\simeq& 
   -\frac{3}{8\pi^2} A_0 \hat{y}_{\ell \alpha} 
   \left(y_\nu^{\dag}L\, y_\nu\right)_{\alpha \beta}
   ,\label{eq:6}\ \ \ 
\end{eqnarray}
where $\Delta M_{LL \alpha \beta}^{e 2}$ and
$\Delta M_{RR \alpha \beta}^{e 2}$ denote the $3 \times 3$ left- and right-handed
slepton soft mass matrices and $\Delta M_{LR \alpha \beta}^{e}$ is the $3 \times 3$ left-right 
mixing slepton matrix that includes a contribution from the 
trilinear couplings. 
Notice, that, as we can see in these equations, at one-loop, flavour
mixing is not induced in the right-handed slepton mass matrix. However,
right-handed slepton mixing is effectively generated from left-handed slepton mixing
in the presence of left-right mixing. Therefore, the mixing in
right-handed sleptons is suppressed by the left-right mixing.
Similarly, flavour mixing is generated in the right-handed sector by 
two-loop effects~\cite{Ibarra:2008uv} but this two-loop induced flavour mixing 
is strongly suppressed by loop factors and the charged lepton Yukawa coupling. 
In the following, we neglect the two-loop contributions and consider only the induced
flavour mixing by the one-loop RGE. Thus, the slepton mass matrix at
the electroweak scale includes only the LFV entries described in
Eqs.~(\ref{eq:4},\ref{eq:5},\ref{eq:6}) that are a function of the neutrino Yukawa coupling.

Moreover, in this paper, we consider a case of the CMSSM where the mass difference between 
the lightest slepton $\tilde l_1$ and the lightest neutralino $\tilde \chi_1^0$, 
$\delta m = m_{\tilde l_1} -  m_{\chi_1^0}$, is smaller than the tau 
mass, $m_\tau$. In the CMSSM with this small mass difference, the LSP is the 
lightest neutralino which is almost pure Bino and the NLSP is the lightest 
slepton which mainly consists of the right-handed 
stau $\tilde \tau_R$. 
The LSP and the NLSP remain the neutralino and the slepton in the 
Seesaw mechanism because the neutrino Yukawa coupling and the Majorana 
mass term do not change mass spectrum significantly. 

In the next section we will discuss the phenomenology of NLSP decays and radiative tau decays 
in this scenario.

\section{NLSP decays and LFV tau branching ratios}
\label{sec:long-lived-slepton}

In this section, we show the decay rates of flavour violating decays of the 
lightest slepton and tau radiative decays. We derive explicit 
relations between the lifetime of the 
lightest slepton and the branching rations of radiative LFV tau decays, $\tau \rightarrow e + \gamma$ and 
$\tau \rightarrow \mu + \gamma$.

\begin{figure}[t]
\begin{center}
\begin{tabular}{ccc}
 \includegraphics[height=39mm]{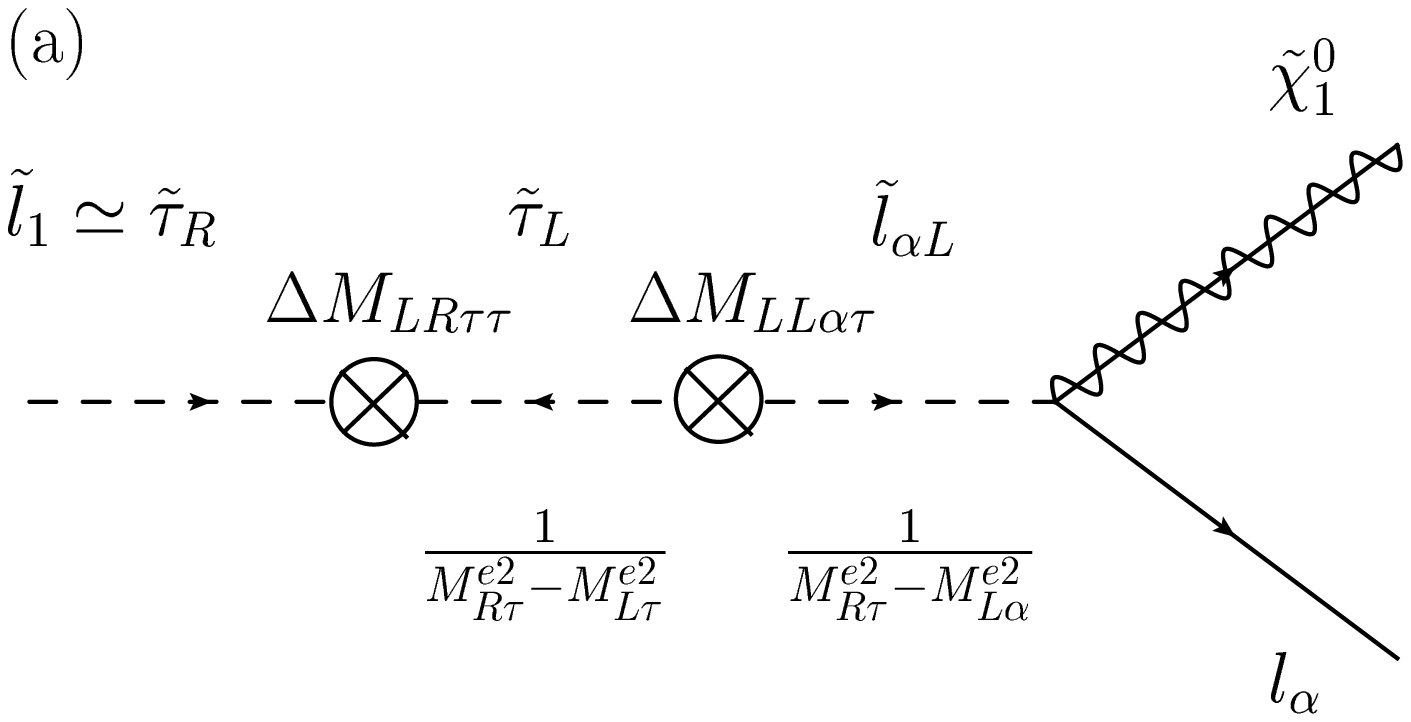} & 
 ~~~ &
 \includegraphics[height=39mm]{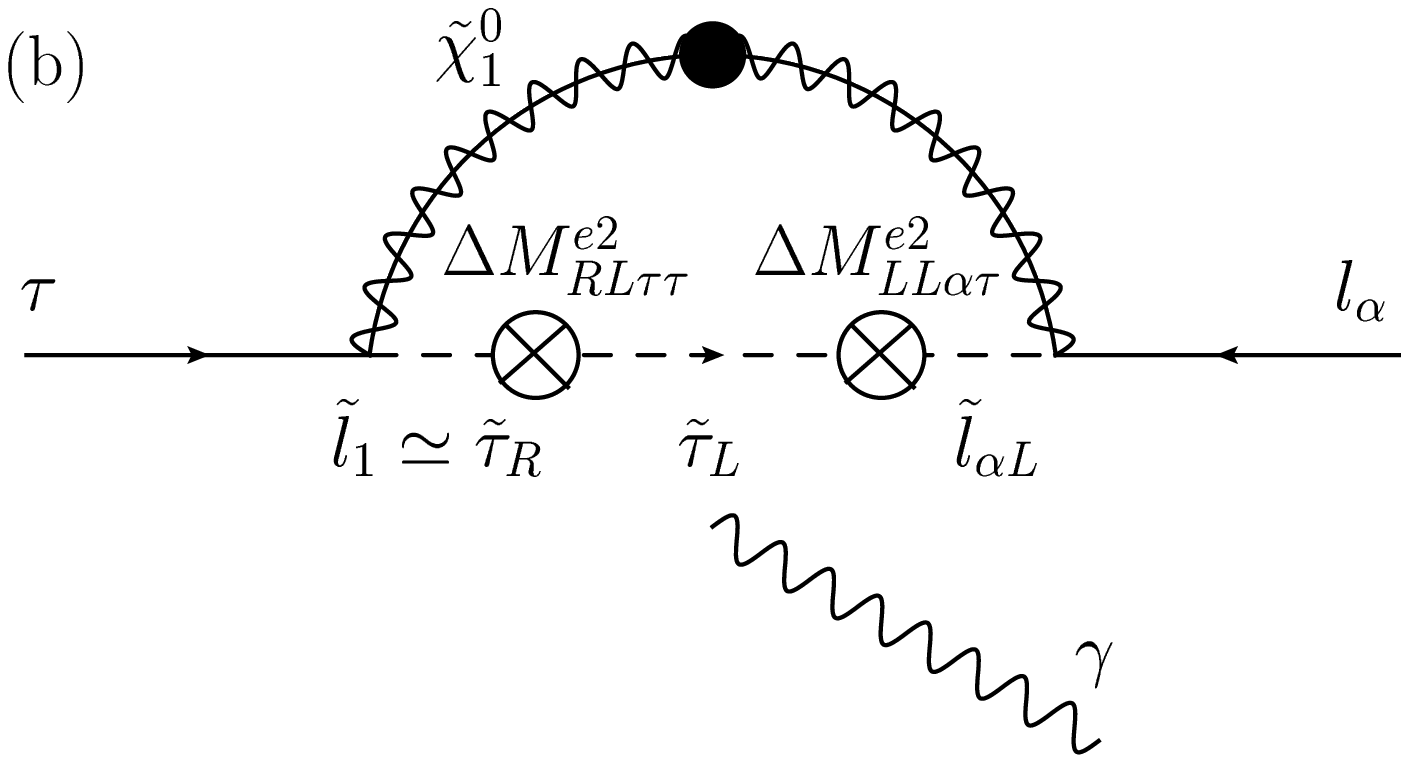}  
\end{tabular}
\end{center}
\caption{Mass insertion Feynman diagrams. Figure (a) depicts the 
2-body decays of the lightest slepton in the presence of 
$\delta^e_{LL}$. Figure (b) depicts radiative LFV tau decays,
$\tau \to l_\alpha 
+ \gamma$, where $\alpha = e$ or $\mu$, from  the neutralino and 
the slepton loop contribution.}
\label{fig:MI_diagram}
\end{figure}

First, we show the decay rates of LFV decays of the lightest slepton. 
The main decay modes of 
the lightest slepton when the mass difference is $m_e~(m_\mu) < \delta m < m_\tau$, 
are LFV 2-body decays, $\tilde{l}_1 \to e~(\mu) + \tilde{\chi}_1^0$, 
unless the induced LFV entries in the slepton mass matrix are so small. In these conditions, the slepton
decay rates are given by \cite{Kaneko:2008re},  
\begin{align}
   \Gamma(\tilde{l}_1 \to l_\alpha + \tilde{\chi}_1^0) 
   &= \frac{g^2_2}{4\pi m_{\tilde{l}_1}} (\delta m)^2 
   |g_{1\alpha 1}^L|^2,\quad~(\alpha = e,~\mu), \label{lfv_2_body}
\end{align}
where $(l_e, l_\mu) = (e,~\mu)$ and $g_2$ is the $SU(2)$ coupling constant. $g^{L}_{1 \alpha 1}$ is 
an effective LFV coupling constant of the neutralino LSP,
the lightest slepton and the charged lepton interaction. 
$g^{L}_{1 \alpha 1}$ can be estimated using 
the Mass Insertion (MI) approximation \cite{Gabbiani:1996hi} as shown Fig.~\ref{fig:MI_diagram}.(a),
\begin{align}
   g^L_{1\alpha 1} 
   \simeq \frac{1}{2}\tan\theta_W  
   \frac{\Delta {M_{LR}^{e 2}}_{\tau\tau}}{ M^{e 2}_{L \tau} - M^{e 2}_{R \tau} } 
   \frac{M^e_{L \alpha} M^e_{R\tau}}{M^{e 2}_{R\tau} - M^{e 2}_{L\alpha}}
   (\delta^e_{LL})_{\alpha \tau}, \label{eL_g} 
\end{align}
where $\theta_W$ is the Weinberg angle, $M^e_{L/R \alpha}$ are diagonal 
elements of the slepton mass matrix and $\Delta {M^{e 2}_{LR}}_{\tau\tau} = m_\tau (A_0 - \mu \tan\beta)$ is 
the flavour-diagonal element of the left-right slepton 
mass matrix. The left-left MI, $(\delta^e_{LL})_{\alpha \tau}$, is defined as 
follows and is generated through RGE running by the neutrino Yukawa coupling,
\begin{eqnarray}
   (\delta_{LL}^e)_{\alpha\beta}
   &=&
   \frac{\Delta M_{LL \alpha \beta}^{e 2}}
   {M^e_{L\alpha} M^e_{L\beta}}
   \simeq
   - \frac{\left(3m_0^2+A_0^2\right) 
   \left(y_\nu^{\dag}L\, y_\nu\right)_{\alpha\beta}}
   {8\pi^2M^e_{L\alpha} M^e_{L \beta}}. 
   \label{eq:2} 
\end{eqnarray}
Then the lifetime of the slepton, 
$\tau_{\tilde{l}_1}$, is approximately given by the inverse of sum of decay rates, 
Eq.~(\ref{lfv_2_body}),
\begin{align}
   \tau_{\tilde{l}_1} 
   \simeq \left( \sum_{\alpha = e,\mu} 
   \Gamma(\tilde{l}_1 \to l_\alpha 
   + \tilde{\chi}^0_1 ) \right)^{-1}. 
   \label{eq:12}
\end{align}
Therefore, the lifetime is inversely proportional to $(\delta m)^2 
(g_{1\alpha1}^L)^2$, and increases as the mass 
difference and/or the LFV couplings become small. 
When the LFV couplings are very small, the sum of 
$\Gamma(\tilde{l}_1 \to l_\alpha + \tilde{\chi}^0_1)$ becomes comparable to or smaller than those 
of flavour conserving 3- or 4-body decay rates, and then  the lifetime becomes 
insensitive to the LFV couplings \cite{Kaneko:2008re}. 
The ratio between the 2-body and 3-body decays is given by~\cite{Jittoh:2005pq}
\begin{equation}
 \begin{split}
    \frac
    {~\Gamma(\tilde l_{1} \to \tilde \chi + \nu_{\tau} + \pi^{\pm})~}
    {\Gamma(\tilde l_{1} \to \tilde \chi_{1}^{0} + e(\mu))} 
    &= 
    \frac
    {~g_{2}^{2} G_{F}^{2} f_{\pi}^{2} 
    \cos^{2}\theta_{c} \tan^{2}\theta_{W} 
    \bigl[ 30 (2 \pi)^{3} m_{\tilde l_{1}} 
    m_{\tau}^{2} \bigr]^{-1} (\delta m)^{6}~}
    {g_{2}^{2} \bigl[ 2 \pi m_{\tilde l_{1}} \bigr]^{-1} 
    (\delta m)^{2} \bigl| g_{1 \alpha 1}^{L} \bigr|^{2}} 
    \\[1mm] & = 
    2.31 \times 10^{-16} \bigl| g_{1 \alpha 1}^{L} \bigr|^{-2} 
    \biggl( \frac{\delta m}{1\text{GeV}} \biggr)^4 , 
 \end{split}      \label{ratio_1}
\end{equation}
and hence Eq.~(\ref{eq:12}) is valid when $g_{1 \alpha 1}^{L} \gsim 
10^{-8}$ for $\delta m = 1$ GeV.  If $\delta m < m_\pi$ ($m_\pi$ 
being the mass of charged pions), the 3-body decay is forbidden. The 
similar ratio with 4-body decay is given by 
\begin{equation}
 \begin{split}
    &\frac{~\Gamma(\tilde l_{1} \to \tilde \chi_{1}^{0} +
    \nu_{\tau} + e + \bar \nu_{e})~}
    {\Gamma(\tilde l_{1} \to \tilde \chi_{1}^{0} + e(\mu))}  \\
    &~~~~~= 
    \frac{~2 g_{2}^{2} G_{F}^{2} \tan^2 \theta_{W} 
    \bigl[ 3 \cdot 5^{3} (2 \pi)^{5} m_{\tilde l_{1}} 
    m_{\tau}^{2} \bigr]^{-1} 
    \delta m \bigl[ (\delta m)^{2} - m_{e}^{2} \bigr]^{5/2} 
    \bigl[ 2(\delta m)^{2} -23 m_{e}^{2} \bigr]~}
    {g_{2}^{2} \bigl[ 2 \pi m_{\tilde l_{1}} \bigr]^{-1}
    (\delta m)^{2} \bigl| g_{1 \alpha 1}^{L} \bigr|^{2}} 
    \\[1mm] &~~~~~ =
    1.80 \times 10^{-24} 
    \bigl| g_{1 \alpha 1}^{L} \bigr|^{-2} 
    \biggl( \frac{\delta m}{50\text{MeV}} \biggr)^6 , 
 \end{split}      \label{ratio_2}
\end{equation}
and hence Eq.~(\ref{eq:12}) is valid when $g_{1 \alpha 1}^{L} \gsim 
10^{-12}$ for $\delta m = 50$ MeV.

Similarly, we can express the decay rates of $\tau \to e + \gamma$ and 
$\tau \to \mu + \gamma$ in terms of the LFV couplings. For simplicity, 
we assume that $\mu\tan\beta$ is large, which is the interesting case
to obtain sizable LFV tau decays. 
In this case, the lightest slepton has a non-negligible mixture with
left-handed stau and the main contribution to LFV tau decays comes
from the pure Bino loop with the lightest slepton\footnote{In some 
parameter space, the Higgsino-Bino mixed loop is also important 
for LFV tau decays (see Fig. 2 in~\cite{Hisano:1995cp}). 
However, in the case that the lightest neutralino is almost pure Bino, 
the mixing is tiny, and as will be shown in numerical results the contribution 
is smaller. Moreover, it has the same dependence on the
flavour-changing MI $(\delta^e_{\rm LL})_{\alpha\beta}$ as the Bino 
loop has, and does not modify the correlation in Eq.~(17). All of contributions from 
neutralinos and sleptons
are included in the numerical analysis in the next section.}. 
Both processes are proportional to $\mu \tan \beta$ and include the 
same MI, $(\delta^e_{LL})_{\alpha \tau}$, that contributes to the 
effective coupling $g^L_{1\alpha 1}$. Therefore, we can expect these 
processes to be strongly correlated, and in fact can estimate the decay 
rates by evaluating Feynmann diagram shown in Fig.~\ref{fig:MI_diagram} 
(b), 
\begin{align}
 \Gamma(\tau \rightarrow l_\alpha + \gamma) 
  \simeq \frac{\pi}{2}
  \left( 
  \frac{\alpha_{em} g_2}{96 \pi^2 \cos\theta_W}
  \right)^2
  \frac{m^3_\tau}{m^2_{\tilde{l}_1}} 
  \big|g^L_{1\alpha 1}\big|^2, \label{eq:8}
\end{align}
where $\alpha_{em}$ is the fine structure constant. The decay rates have 
the same dependence on $g^L_{1\alpha1}$ as Eq.~(\ref{lfv_2_body}).

Using the  above results, we can derive useful relations between the lifetime 
of 
the slepton and the branching ratios of the LFV tau decays, Br($\tau \to 
e~(\mu) + \gamma$). It can be done straight-forwardly by replacing the coupling
$|g^L_{1\alpha 1}|^2$ in Eq.~(\ref{lfv_2_body}) with that in Eq.~(\ref{eq:8}). 
In the case of $m_\mu < \delta m < m_\tau$, the branching ratios are 
expressed
\begin{align}
   \mathrm{Br}(\tau \to e + \gamma) 
   + \mathrm{Br}(\tau \to \mu + \gamma) 
   = 
   2 m_\tau^3 \tau_\tau 
   \left( \frac{\alpha_{em}}{96 \pi \cos\theta_W} \right)^2 
   \bigg(\tau_{\tilde{l}_1} m_{\tilde{l}_1} 
   (\delta m)^2 \bigg)^{-1}.
   \label{eq:9}
\end{align}
From Eq.~(\ref{eq:9}), we can obtain an upper bound on the branching ratios of the radiative tau decay 
when the lifetime and the mass difference are given. For $m_e < \delta m < m_\mu$, 
the lifetime is related only to Br($\tau \to e + \gamma$),
\begin{align}
   \mathrm{Br}(\tau \to e + \gamma) 
   = 
   2 m_\tau^3 \tau_\tau 
   \left( \frac{\alpha_{em}}{96 \pi \cos\theta_W} \right)^2 
   \bigg(\tau_{\tilde{l}_1} m_{\tilde{l}_1} 
   (\delta m)^2 \bigg)^{-1}, 
   \label{eq:15}
\end{align}
because the slepton decay into $\mu + \tilde{\chi}^0_1$ is kinematically 
forbidden for this mass difference. When the 2-body slepton decays are dominant, 
the relations, Eqs.~(\ref{eq:9}) and 
(\ref{eq:15}), can be expressed in a different way,
\begin{align}
   \mathrm{Br}(\tau \to l_\alpha + \gamma) 
   = 2 m_\tau^3 \tau_\tau 
   \left( \frac{\alpha_{em}}{96 \pi \cos\theta_W} \right)^2 
   \bigg(\tau_{\tilde{l}_1} m_{\tilde{l}_1} 
   (\delta m)^2 \bigg)^{-1}
   \mathrm{Br}(\tilde{l}_1 \to l_\alpha + \tilde{\chi}^0_1), 
   \label{eq:13}
\end{align}
where Br($\tilde{l}_1 \to l_\alpha + \tilde{\chi}^0_1$) is the branching 
ratio of the slepton decay.
From Eq.\eqref{eq:13},  the branching ratios of LFV tau decays 
can be predicted once the mass difference, the branching ratios 
and the lifetime of the slepton are determined.

Another relation obtained from Eqs.~(\ref{lfv_2_body}) and (\ref{eq:8}) 
is 
\begin{align}
   \frac{\mathrm{Br}(\tau \to e + \gamma)}
   {\mathrm{Br}(\tau \to \mu + \gamma)}
   =
   \frac{\mathrm{Br}(\tilde{l}_1 \to e + \tilde{\chi}_1^0)}
   {\mathrm{Br}(\tilde{l}_1 \to \mu + \tilde{\chi}^0_1)}
   \equiv r_{e\mu}, 
   \label{eq:14}
\end{align}
which is valid only for $m_\mu < \delta m < m_\tau$. The ratio of the 
branching ratios of the tau decays is determined when $r_{e\mu}$ is 
determined through the slepton decays. The ratio can be given in terms 
of the LFV couplings,
\begin{align}
   r_{e\mu} 
   \simeq 
   \frac{(g_{1 e 1}^L)^2}
   {(g_{1 \mu 1}^L)^2}.
\end{align}
This ratio can be predicted once the LFV couplings are given in a specific 
model. It is important to emphasize here that these relations hold 
also for models where the LFV entries in the slepton mass matrix are not generated by the RGEs in a Seesaw scenario. Both the lifetime and the branching 
ratios depend on these LFV parameters and these expressions are still valid 
in other scenarios if $\Delta 
M_{LL \alpha \beta}^{e 2} \gg \Delta M_{RR \alpha \beta}^{e 2}$ 
and $\mu \tan \beta \gg m_{\mathrm{SUSY}}$, 
$m_{\mathrm{SUSY}}$ being a typical mass scale of particles in the loop in Fig.~\ref{fig:MI_diagram}

\section{Numerical analysis}
\label{sec:numerical-result}

In this section we analyze numerically the lightest slepton lifetime and the radiative tau decays in the CMSSM in the 
parameter region corresponding
to nearly degenerate slepton and neutralino as described above.
All these results on the lifetime of the slepton and the branching ratios of 
$\tau \rightarrow e~(\mu) + \gamma$ and also $\mu \rightarrow e + \gamma$ 
are obtained by solving one-loop RGEs and using exact formulae of 
the decay rates given in \cite{Hisano:1995cp,Kaneko:2008re}.

Regarding the CMSSM parameters, 
we choose three reference points shown in Table.~\ref{tab:mSUGRApoints} 
to see difference of the lifetimes and the branching ratios on 
$\delta m$.
The point $1$ corresponds to the case where $\delta m$ is slightly smaller than $m_\tau$ 
while the point $3$ is the one where $\delta m$ is smaller than $m_\mu$, 
the point $2$ is an intermediate value between the point $1$ and $3$. The masses
of the NLSP slepton and the LSP neutralino are almost the same for three points 
and are approximately $360$ GeV. 
It is shown in \cite{Kaneko:2008re} that, in the CMSSM parameter space, 
there exist regions for $\delta m < m_\tau$ where not only the dark matter 
abundance but also the Higgs mass, the deviation of the muon 
anomalous magnetic moment and the branching ratios of $b \to s + \gamma$ 
are consistent with experimental bounds. The points we choose here are 
included in this region. 
\begin{table}[t]
\begin{center}
 \begin{tabular}{|c|c|c|c|} \hline
  & ~~~~point $1$~~~~ 
  & ~~~~point $2$~~~~ 
  & ~~~~point $3$~~~~ 
  \\ \hline
  ~~~~~$m_0$ (GeV)~~~~~       
  & $325.1$ 
  & $275.9$ 
  & $351.9$ 
  \\ \hline
  $M_{1/2}$ (GeV)             
  & $859.7$ 
  & $870.6$ 
  & $821.0$ 
  \\ \hline
  $A_0$ (GeV)                 
  & $664.0$ 
  & $161.2$ 
  & $857.8$ 
  \\ \hline
  $\tan\beta$                 
  & $36.68$ 
  & $31.88$ 
  & $38.23$ 
  \\ \hline
  $\delta m$  (GeV)           
  & $1.44$ 
  & $0.212$  
  & $0.071$  
  \\ \hline
 \end{tabular}
\end{center} 
\caption{Reference points of the CMSSM parameters. 
$\delta m$ is the mass difference between the 
lightest slepton and the lightest neutralino in 
absence of flavour mixing.}
\label{tab:mSUGRApoints}
\end{table}

In this work, flavour mixing is generated by the neutrino Yukawa
coupling through RGE evolution. Therefore, we need to specify the
neutrino Yukawa coupling consistent with the observed neutrino masses
and mixing, Eq.~(\ref{eq:7}). Mixing angles of the left-handed
neutrinos ,$\theta_{12},\theta_{23},\theta_{13}$, have been measured by neutrino oscillation experiments
\cite{Fukuda:2002pe, Ahmed:2003kj,Araki:2004mb,Ahn:2006zza}. We use
following values \cite{Amsler:2008zzb},
\begin{align}
  \tan^2 \theta_{12} &= 0.45, 
  \qquad \sin^2 2\theta_{23} = 1.0, 
\end{align}
and we assume that $\theta_{13}$ and the CP violating phases are zero. 
For the left-handed neutrino masses, we employ the normal hierarchy, 
$m_2 < m_3$ and $m_1 = 0$, for simplicity. Then $m_2$ and $m_3$ 
are determined by two squared mass differences, $\Delta m^2_{ij} 
\equiv m_i^2 - m_j^2$,
\begin{align}
   \Delta m^2_{21} 
   &
   = 8.0 \times 10^{-5} \mathrm{eV}^2, \qquad \Delta m^2_{31}
   = 2.0 \times 10^{-3} \mathrm{eV}^2.
\end{align}
The undetermined parameters in Eq.~(\ref{eq:7}) are the Majorana masses, 
$\hat{M}_R$, and 
the complex angles, $\omega_{1,2,3}$. These variables can not be 
directly determined by near future experiments, perhaps in the future, and 
could only be indirectly inferred through LFV experiments in a scenario of 
supersymmetric seesaw mechanism. Thus, for us, they are free parameters 
in our model. 

As mentioned in Sec.~\ref{sec:setup:-mssm-with-RN}, the flavour violating
entries on the slepton mass matrix are strongly dependent on the complex
angles $\omega_{1,2,3}$. In this work, we do not intend to do a full analysis of the 
possible neutrino Yukawa coupling and only present some representative examples. 
 Hence, for simplicity, we fix the Majorana masses assuming a normal hierarchy
\footnote{In fact, we performed a similar analysis taking
  $M_{R3} = 10^{14}$ GeV and we did not find significant
  differences. Moreover, the assumption of the normal hierarchy for the
  left-handed and the right-handed neutrino masses is not critical in
  our discussion. A different hierarchy could give different numerical
  results for the studied lifetimes and branching ratios but would not
  change the correlations studying here.},
\begin{align}
 M_{R1} & = 10^{10}~~\mathrm{GeV}, \\
 M_{R2} & = 10^{11}~~\mathrm{GeV}, \\
 M_{R3} & = 10^{12}~~\mathrm{GeV},
\end{align}
and vary $\omega_{1,2,3}$ in the following ranges, 
\begin{align}
   0 & \le \mathrm{Re}(\omega_1), 
   ~\mathrm{Re}(\omega_2),
   ~\mathrm{Re}(\omega_3) \le 2\pi, 
   \\
   -\frac{3}{2}\pi & \le \mathrm{Im}(\omega_1), 
   ~\mathrm{Im}(\omega_2), 
   ~\mathrm{Im}(\omega_3) \le \frac{3}{2}\pi.
\end{align}
Strictly speaking, the imaginary parts of $\omega_{1,2,3}$ can be taken from
$-\infty$ to $+\infty$. However, as can be seen from Eq.~(\ref{eq:7}), if
we take large values of the imaginary parts, the induced flavour
mixing will easily exceed the present experimental bounds.  We have
checked that, if we restrict Im$(\omega_{1,2,3})$ to the above range, the
obtained branching ratios do not exceed the present bounds by a large amount.


\begin{figure}[t]
\begin{center}
\begin{tabular}{cc}
 \includegraphics[width=84mm]{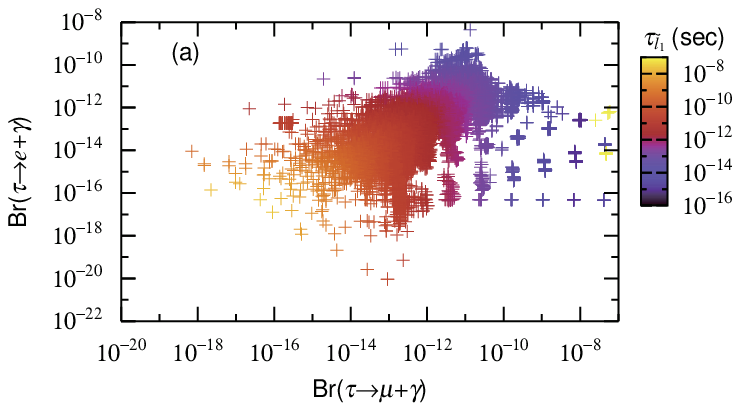} 
  &
 \includegraphics[width=80mm]{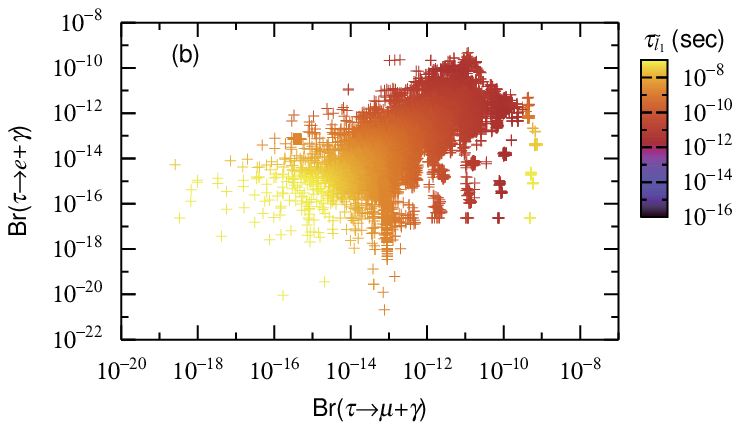} 
  \\
  & \\
 \includegraphics[width=80mm]{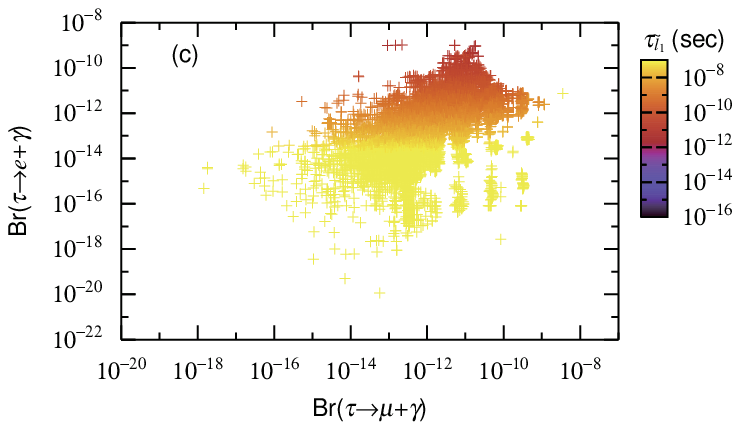} 
\end{tabular}
\end{center}
\caption{Scatter plots of the lifetime of the 
slepton as a function of Br($\tau \to \mu + 
\gamma$) and Br($\tau \rightarrow e + 
\gamma$) for three reference points. 
Values of the lifetime are indicated at colour bar 
on the side of figures. Figures (a), (b) and (c) 
correspond to the point $1$, $2$ and $3$.}
\label{fig:BrTauMG_BrTauEG_lifetimes}
\end{figure}
In Fig.~\ref{fig:BrTauMG_BrTauEG_lifetimes}, we show the lightest slepton 
lifetime, $\tau_{\tilde{l}_1}$,  in different colours as a function of the
branching ratios Br$(\tau \to e + \gamma)$ and Br$(\tau \rightarrow \mu + \gamma)$ as a scatter plot varying the complex 
angles in the previously defined. 
Panels (a), (b) and (c) correspond to the CMSSM points $1$, $2$ and $3$, 
respectively. In all panels, the experimental bound on the branching 
ratio Br$(\mu \to e + \gamma) < 1.2 \times 10^{-11}$ 
\cite{Amsler:2008zzb}, is imposed and it excludes the right-top 
corner (white region) on each panel. This is because, in this region, 
both $g_{1e1}^L$ and $g_{1 \mu 1}^L$, or both the corresponding mass insertions $(\delta_{LL}^e)_{e \tau}$ 
and $(\delta_{LL}^e)_{\mu \tau}$ are large. Then,  
even if $(\delta_{LL}^e)_{e \mu}$ is small enough, 
sizable $\mu \to e + \gamma$ occurs via a double mass insertion which picks up 
$(\delta_{LL}^e)_{e \tau}$ and $(\delta_{LL}^e)_{\mu \tau}$ on the slepton line in the loop.
We can also see, in this figure, that the slepton lifetime in the points 
$1$ and $2$ are strongly correlated 
with Br($\tau \to e + \gamma$) $+$ Br($\tau \to \mu + \gamma$) 
as shown in Eq.~(\ref{eq:9}), while the lifetime in the point $3$ depends on only 
Br$(\tau \rightarrow e + \gamma)$ as shown in Eq.~(\ref{eq:15}). 
This is due to the fact that the slepton can decay 
into $e$ or $\mu$ with the neutralino for $m_\mu < \delta m < m_\tau$. 
Therefore the lifetime depends on both of $g_{1e1}^L$ and $g_{1 \mu 1}^L$ 
on which the branching ratios, Br$(\tau \to e + \gamma)$ and Br$(\tau \to \mu + \gamma)$, 
also depend. On the other hand, the slepton can decay 
only into $e$ and the neutralino for $m_e < \delta m < m_\mu$. Therefore 
the lifetime depends on only $g_{1e1}^L$ and is correlated with only Br$(\tau \to e + \gamma)$.
In Figs.~\ref{fig:BrTauMG_BrTauEG_lifetimes}.(a) and (b), for the same
values of the branching ratios, the lifetimes become longer as the
mass differences become smaller. This is because the lifetime is
inversely proportional to $(\delta m)^2$ as shown in
Eq.~\eqref{lfv_2_body}.

From Eq.~(\ref{eq:9}), we can see that the experimental upper bound  on the branching ratios 
implies a lower bound on the lightest slepton lifetime. Due to looser experimental bounds on the branching ratios of 
the radiative tau decays, the branching ratio of $\mu \to e + \gamma$ provides the lower bound on the lifetime 
through the double mass insertions.
\begin{table}[t]
\begin{center}
 \begin{tabular}{|c|c|c|c|} \hline
   & ~~~point $1$~~~  
   & ~~~point $2$~~~ 
   & ~~~point $3$~~~ 
   \\ \hline\hline
   ~~upper bound~(sec.)~~ 
   &~~~$1.05 \times 10^{-7}$~~~
   &~~~$2.85 \times 10^{-1}$~~~
   &~~~$3.53 \times 10^4$~~~ 
   \\ \hline
   ~~lower bound~(sec.)~~ 
   &~~~$7.98 \times 10^{-16}$~~~
   &~~~$3.13 \times 10^{-12}$~~~
   &~~~$4.22 \times 10^{-12}$~~~ 
   \\ \hline
 \end{tabular}
\end{center} 
\caption{
Upper and lower bounds on the lifetime 
of the stau. Upper bounds are calculated 
from flavour conserving 3- and 4-body 
decays rates given in \cite{Jittoh:2005pq}.}
\label{tab:lifetime_bounds}
\end{table}
On the other hand, as mentioned 
in Sec.~\ref{sec:long-lived-slepton}, for very small LFV couplings, the lifetime becomes insensitive to  
$g_{1 \alpha 1}^{L}$ for $g_{1 \alpha 1}^{L} \lsim 10^{-8}$, Eq.~\eqref{ratio_1}, and the 
lifetime is given by decay rates of the flavour conserving 3-body and/or 4-decays. In the case of $\delta m < m_\pi$, the lifetime 
is constant for $g_{1 \alpha 1}^{L} \lsim 10^{-12}$, Eq.~\eqref{ratio_2}, and is determined by decay 
rate of the flavour conserving 4-body decays. 
The bounds are summarized in Table~\ref{tab:lifetime_bounds}. 
One can see from this Table~\ref{tab:lifetime_bounds} that, in the degenerate slepton-neutralino 
region, the lifetime of the slepton can change in many orders of magnitude in the presence of LFV 
couplings. The off-diagonal entries in the
left-left part of the slepton mass matrix, Eq.~\eqref{eq:4}, depends
on the complex orthogonal matrix, $W$, through the Yukawa coupling of 
neutrinos. Thus, the lifetime depends on $W$ and change in 
many orders of magnitude by varying the imaginary parts of $\omega_{1,2,3}$. 
It is important to emphasize here that, as was discussed in 
\cite{Kaneko:2008re}, the ATLAS detector could measure the lifetimes of 
the slepton and therefore determine or constrain the LFV couplings.
Thus, in the case of the small $\delta m$, the LHC experiment can provide an 
opportunity to obtain information on the LFV parameters and hence on the parameters in the Seesaw mechanism.

\begin{figure}[t]
\begin{center}
\begin{tabular}{cc}
 \includegraphics[width=82mm]{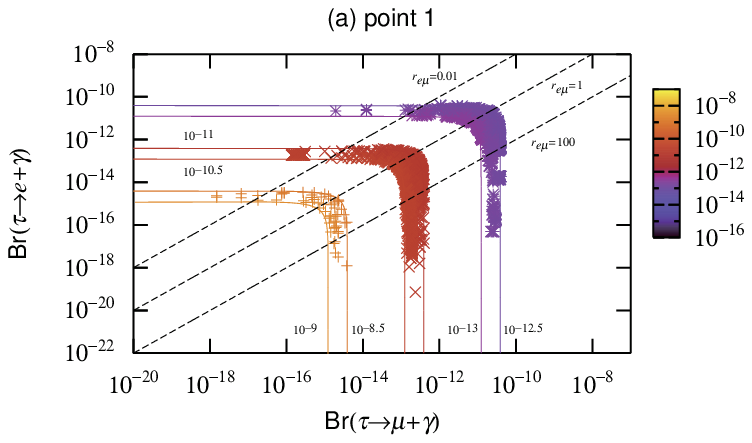} 
&
 \includegraphics[width=82mm]{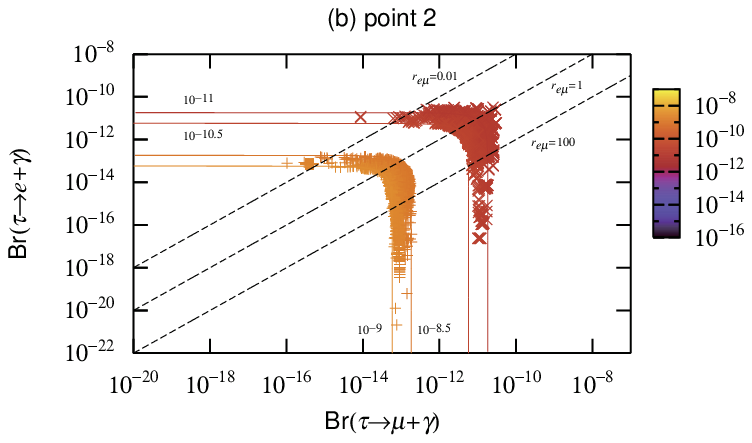} 
\end{tabular}
\end{center}
\caption{
Contour plots of branching ratios of $\tau 
\to \mu + \gamma$ and  $\tau \to e + 
\gamma$, and the lifetime of the slepton for 
the points $1$ and $2$. The lifetime of $10^{-9} 
- 10^{-8.5}$, $10^{-11} - 10^{-10.5}$ 
and $10^{-13} - 10^{-12.5}$ sec. are 
shown. Solid curves are drawn by 
Eq.~(\ref{eq:9}) and corresponding 
lifetimes are shown near the curves. Dashed 
lines are drawn by Eq.~(\ref{eq:14}) and values of 
$r_{e\mu}$ are shown near the lines.}
\label{fig:BrTauMG_BrTauEG_lifetimes_separate}
\end{figure}

\begin{table}[t]
\begin{center}
 \begin{tabular}{|c|c|c|} \hline
   & point $1$ 
   & point $2$ 
   \\ \hline\hline
   ~~~~$\tau_{\tilde{l}_1}$~~~~ 
   & ~~~~Br$(\tau \to e~(\mu) + \gamma)$~~~~ 
   & ~~~~Br$(\tau \to e~(\mu) + \gamma)$~~~~ 
   \\ \hline
   ~~~$10^{-9} - 10^{-8.5}$ ~~(sec.)~ 
   & $(1.23 - 3.90) \times 10^{-15}$ 
   & $(0.580 - 1.84) \times 10^{-13}$ 
   \\ \hline
   ~~~~$10^{-11} - 10^{-10.5}$ (sec.)~~
   & $(1.23 - 3.90) \times 10^{-13}$ 
   & $(0.580 - 1.84) \times 10^{-11}$ 
   \\ \hline
   ~~~~$10^{-13} - 10^{-12.5}$ (sec.)~~
   & $(1.23 - 3.90) \times 10^{-11}$ 
   & --- \\ \hline
 \end{tabular}
\end{center} 
\caption{Upper bounds on Br($\tau \to e~(\mu) 
+ \gamma$) for the points $1$ and $2$.}
\label{tab:Br_bounds}
\end{table} 
In Figs.~\ref{fig:BrTauMG_BrTauEG_lifetimes_separate} (a) and (b) we
plot the branching ratios Br($\tau \to e~(\mu) + \gamma$) for the points
$1$ and $2$ selecting values of the slepton lifetimes in the ranges,
$\tau_{\tilde{l}_1}= 10^{-9} - 10^{-8.5}$, $10^{-11} - 10^{-10.5}$ and
$10^{-13} - 10^{-12.5}$ sec. All the points in this figure are
extracted from Fig.~\ref{fig:BrTauMG_BrTauEG_lifetimes} (a) and (b).
The solid curves in these panels represent the correlation shown in
Eq.~(\ref{eq:9}). We can see that Eq.~(\ref{eq:9}) is in very good
agreement with the numerical results for the points $1$ and $2$, and 
the branching ratios are predicted once the lifetime, the mass of the
slepton and the mass difference are given.
These correlations between the LFV branching ratios and the lightest slepton 
lifetime are summarized in Table.~\ref{tab:Br_bounds} for both points $1$ and 
$2$.
In fact, the LFV branching ratios can be completely determined from slepton decays using  $r_{e\mu}$ 
which is determined by measuring the slepton decays to the different leptonic flavours. 
In this figure, dashed lines represent Eq.~(\ref{eq:14}) 
and the corresponding values of $r_{e\mu}$ are shown near the lines. 
A large value of 
$r_{e\mu}$ corresponds to larger branching ratio of 
$\tau \to e + \gamma$ than that of $\tau \to \mu + \gamma$. 
For example, for $r_{e\mu} = 100$ and the lifetime 
between $10^{-8.5} - 10^{-9}$ sec. in 
Fig.~\ref{fig:BrTauMG_BrTauEG_lifetimes_separate} (a), Br$(\tau \to e + 
\gamma)$ is roughly between $10^{-17}$ and $5 \times 10^{-17}$ and 
Br$(\tau \to \mu + \gamma)$ is between $10^{-15}$ and $5 \times 
10^{-15}$.

\begin{figure}[t]
\begin{center}
\begin{tabular}{cc}
 \includegraphics[width=80mm]{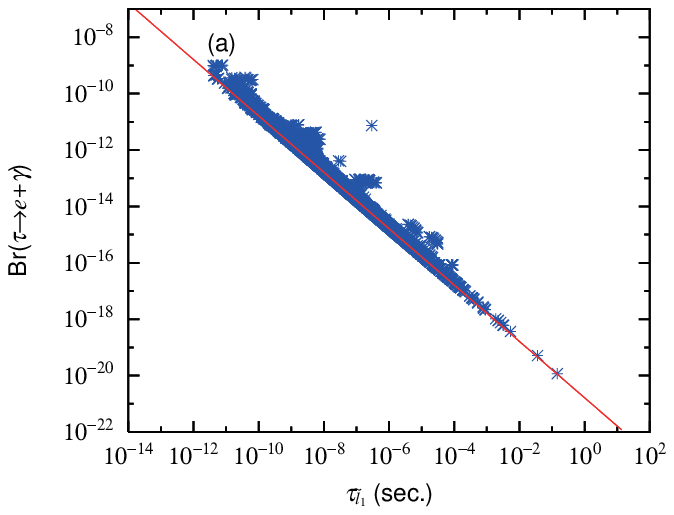} 
&
 \includegraphics[width=80mm]{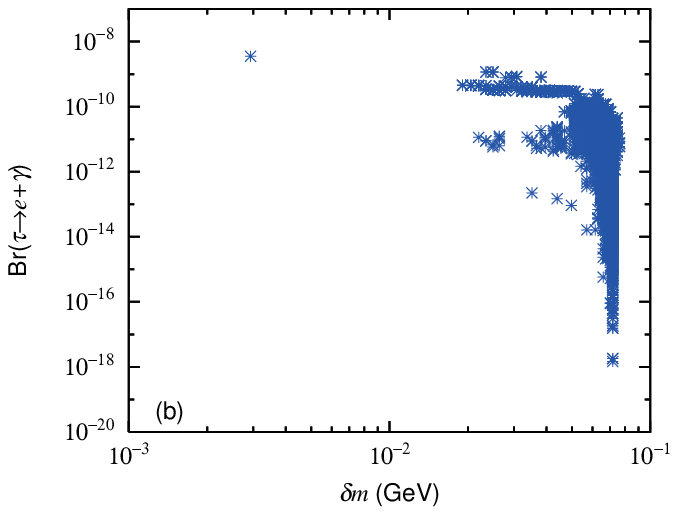} 
\end{tabular}
\end{center}
\caption{Plots of the branching ratio of $\tau \to 
e + \gamma$ as a function of the slepton lifetime, 
$\tau_{\tilde{l}_1}$, (a) and as a function of 
$\delta m$, (b) for the point $3$. Solid line in 
the left panel, (a), shows Eq.~(\ref{eq:15}).}
\label{fig:BrTauEG_lifetime_dm}
\end{figure}

Similarly, Fig.~\ref{fig:BrTauEG_lifetime_dm} (a) shows Br$(\tau \to e + \gamma)$ 
as a function of the slepton lifetime for the point $3$. Now, the solid line represents 
Eq.~(\ref{eq:15}) with $\delta m = 0.071$ GeV. We can see that Br$(\tau \to 
e + \gamma)$ is inversely proportional to the lifetime and Eq.~(\ref{eq:15}) 
is in very good agreement with the numerical analysis. 
In this case, we can directly determine Br$(\tau \to e + \gamma)$ by measuring 
the lightest slepton lifetime. For instance, a lifetime of $10^{-8.5}$ corresponds to a branching ratio is $4.56 \times 10^{-13}$ and for a lifetime of 
$10^{-10.5}$, the branching ratio is $4.36 \times 10^{-11}$. 
We also see in this Figure that there are some points which deviate from 
Eq.~(\ref{eq:15}). This deviation stems from changes of the mass 
difference. If the neutrino Yukawa coupling is large, relatively large flavour 
mixing in the slepton mass matrix is induced through RGE and this reduces the mass of the 
lightest slepton. In Fig.~\ref{fig:BrTauEG_lifetime_dm} (b), we show 
Br$(\tau \to e+ \gamma)$ in terms of the mass difference. 
It can be seen here, that starting from $\delta m = 0.071$ GeV in the
absence of LFV mass insertions, the mass difference can become much
smaller when $(\delta_{LL}^e)_{e \tau}$, i.e. Br$(\tau \to e +
\gamma)$, is large. Note that for the point $1$ and $2$, the change
on the slepton mass is negligible because the mass difference is not
as small as that for the point $3$.

\section{Summary and Discussion}
\label{sec:conclusion}

In this work, we have studied several lepton flavour violation observables 
in the supersymmetric Seesaw scenario in the region of nearly degenerate 
slepton and neutralino. Within the Seesaw scenario, flavour violating 
entries in the slepton mass matrix are 
induced only in the left-left and left-right parts at one-loop order, and LFV is 
transfered to the right-handed sleptons via left-right part. 
When the mass difference between the NLSP, the lightest slepton, 
and the LSP, the lightest neutralino, is smaller than the tau mass, the 
lifetime of the lightest slepton is inversely proportional to the LFV entries
in the slepton mass matrix.

We have seen in Sec.~\ref{sec:long-lived-slepton}, that, in the case of 
$\delta m < m_\tau$, flavour violating 2-body decay rates of the lightest 
slepton are  proportional to the square of the effective couplings 
$g_{1e(\mu)1}^{L}$.
We have also shown that the branching ratios of $\tau \to e~(\mu) + 
\gamma$ are proportional to the same LFV couplings. Then, using these 
results, we derived relations between the lifetime of the slepton and the 
branching ratios of the LFV tau decays. 
We also found that the ratio of Br($\tau \to \mu + \gamma$) to 
Br($\tau \to e + \gamma$) is the same as that of Br($\tilde{l}_1 \to \mu 
+ \tilde{\chi}^0_1$) to Br($\tilde{l}_1 \to e + \tilde{\chi}^0_1$). 
Using these relations, it is possible to determine the branching ratios of 
LFV tau decays by measuring the lifetime and the branching ratios of the 
LFV slepton decays.  
Note that these relations are valid in any scenarios with $\delta m < m_\tau$ 
when $(\Delta M^{e 2}_{LL})_{\alpha \tau} \gg 
(\Delta M^{e 2}_{RR})_{\alpha \tau}$ and $\mu\tan \beta \gg m_{\mathrm{SUSY}}$ are satisfied.

Then, in Sec.~\ref{sec:numerical-result}, we have checked these relations 
numerically by calculating the NLSP lifetime 
and tau LFV branching ratios by varying the complex phases in $W$. We chose three 
reference points of the CMSSM parameters which correspond to $\delta 
m = 1.44,~0.212$ and $0.071$ GeV, respectively. 
As summarized in Table \ref{tab:lifetime_bounds}, the lifetime of the
lightest slepton, in this scenario, can be reduced in many orders of
magnitude when compared with the flavour conserving case due to the
LFV entries. These LFV entries are determined by the complex
mixings, $\omega_{1,2,3}$, which can not be determined by low-energy
experiments. Fixing the right-handed Majorana masses and varying the
complex angles $\omega_{1,2,3}$, the slepton lifetimes can be found
between $7.98 \times 10^{-16}$ and $1.05 \times 10^{-7}$ sec. for
$\delta m = 1.44$ GeV, between $3.13 \times 10^{-12}$ and $2.85 \times
10^{-1}$ sec. for $\delta m = 0.212$ GeV, and between $4.22 \times
10^{-12}$ and $3.53 \times 10^4$ sec. for $\delta m = 0.071$ GeV. As
shown in Ref.~\cite{Kaneko:2008re}, the ATLAS detector will be able to
determine the lifetime between $10^{-12}$ and $10^{-5}$ sec. Therefore,
it would be possible to measure very small LFV entries through
lightest slepton decays at the LHC experiments.
In this way, slepton lifetimes for mass differences 
$m_\mu < \delta m < m_\tau$ are 
related with both Br($\tau \to e + \gamma$) and Br($\tau \to \mu + 
\gamma$), while lifetimes for $m_e < \delta m < m_\mu$ are related with 
only Br($\tau \to e + \gamma$). 
Thus, by measuring at the LHC the lightest slepton lifetime, the
branching ratios to the $e$ and $\mu$ channel and the mass difference,
we can predict the branching ratios of the radiative tau decays in the slepton-neutralino coannihilation 
scenario with the supersymmetric Seesaw mechanism. 

\acknowledgments    

The work of J. S. was supported in part by the Grant-in-Aid for the Ministry of Education, Culture, 
Sports, Science, and Technology, Government of Japan Contact No. 20540251. 
The work of T.~S was supported in part by the Grant-in-Aid for the Ministry of
Education, Culture, Sports, Science, and Technology, Government of Japan Contact 
No. 19540284.
The work of T.~S. and O.~V. was supported in part by MEC and FEDER (EC), Grant No. FPA2008-02878 and by
the Generalitat Valenciana under the grants PROMETEO/2008/004 and 
GVPRE/2008/003. 
The work of O.~V was also supported in part by European program MRTN-CT-2006-035482 ``Flavianet''. 
T.~S. is the Yukawa Fellow and his work is supported in part by Yukawa Memorial Foundation.

\bibliographystyle{apsrev}
\bibliography{biblio}

\end{document}